\documentclass[12pt]{article}
\usepackage{epsfig}
\usepackage{amssymb}
\usepackage{citesort}
\usepackage{latexsym}

\newsavebox{\LSIM}
\sbox{\LSIM}{\raisebox{-1ex}{$\ \stackrel{\textstyle<}{\sim}\ $}}
\newcommand{\lsim}{\usebox{\LSIM}}
\newsavebox{\GSIM}
\sbox{\GSIM}{\raisebox{-1ex}{$\ \stackrel{\textstyle>}{\sim}\ $}}

\begin{document}
\begin{titlepage}
\begin{flushright}
BI-TP 2005/14\\
hep-ph/0504222
\end{flushright}
$\mbox{ }$
\vspace{.1cm}
\begin{center}
\vspace{.5cm}
{\bf\Large Electroweak Baryogenesis with dimension-6}\\[.3cm]
{\bf\Large Higgs interactions}\footnote{Talk given at
  the 8th INTERNATIONAL MOSCOW SCHOOL OF PHYSICS}\\
\vspace{1cm}
Lars Fromme\footnote{fromme@physik.uni-bielefeld.de}\\ 
 
\vspace{1cm} {\em  
Fakult\"at f\"ur Physik, Universit\"at Bielefeld, D-33615 Bielefeld, Germany}\\

\end{center}
\bigskip\noindent
\vspace{1.cm}

\begin{abstract}

We present the computation of the baryon asymmetry in the SM amplified by
dimension-6 Higgs interactions using the WKB approximation. Analyzing the
one-loop potential it turns out that the phase transition is strongly first
order in a wide range of the parameters. It is ensured not to wash out the net
baryon number gained previously even for Higgs masses up to at least 170 GeV. In
addition dimension-6 operators induce new sources of CP violation. Novel source
terms which enhance the generated baryon asymmetry emerge in the transport
equations. This model predicts a baryon to entropy ratio close to the observed
value for a large part of the parameter space.\\
My talk was mainly based on our recent work \cite{wir}.\\

\end{abstract}
\end{titlepage}
\section{Introduction}
The baryon asymmetry of the universe has been measured by combining measurements
of the cosmic microwave background and large scale structures. The baryon to
entropy ratio
\begin{equation} 
\label{eta}
\eta_B\equiv \frac{n_B}{s}=(8.9\pm0.4)\times10^{-11}
\end{equation}
describes the abundance of matter over anti-matter in the universe. The origin
of this value we want to explain.\\
In 1967 Sakharov formulated three necessary ingredients for baryogenesis, these
are: baryon number non-conservation, C and CP violation and deviation from
thermal equilibrium. The first condition is obvious, there must be a mechanism
which violates B, since we assume the universe to be baryon symmetric in the
beginning. Secondly, if C and CP were conserved, particles and anti-particles
were produced with an equal rate so that a net baryon number cannot be
generated. Thirdly we need deviation from equilibrium because CPT invariance
ensures vanishing baryon density in strict thermal equilibrium. For electroweak
baryogenesis a considerable departure from equilibrium is only possible from a
first order phase transition (PT), since the expansion of the universe is slow
during the electroweak epoch. The transition is triggered by an energy barrier
in the Higgs potential which separates two energetically degenerate phases at
the critical temperature $T_c$. At this temperature the formation of bubbles
starts, they expand and finally percolate. The vacuum expectation value of the
Higgs field is zero in the symmetric phase and changes rapidly at the bubble
wall to a non-zero value $v_c$ inside the bubbles. Baryon number violation takes
place outside the bubbles while within the sphaleron induced (B+L)-violating but
(B-L)-conserving reactions must be strongly suppressed. To prevent baryon number
washout after the phase transition the so called "washout criterion"
\begin{equation}\label{wo}
\xi={v_c\over T_c} \ge 1.1
\end{equation}
has to be satisfied \cite{Moore}. This is the condition for a first order
transition to be strong.\\
The baryon asymmetry cannot be explained within the standard model (SM) due to
the facts that there is not enough CP violation and the transition is only a
smooth cross over. To overcome these difficulties we need extensions to the
SM. One possible scenario is to introduce a non-renormalizable $\phi^6$ operator
\cite{Z93,GSW04}. The corresponding potential is
\begin{equation}
V(\phi)=-{\mu^2 \over 2}\phi^2+{\lambda \over 4}\phi^4+{1\over
  {8M^2}}\phi^6,
\end{equation}
where $\phi^2\equiv 2\Phi^{\dagger}\Phi$ with the SM Higgs doublet $\Phi$.
In such a model the barrier in the potential
which triggers the first order transition is no longer only provided by
the cubic one-loop thermal corrections of the weak gauge bosons. It can also be
generated from a negative $\phi^4$ term because the Higgs potential is
stabilized by the $\phi^6$ interaction. This yields to a strong first order
PT for Higgs masses larger than the experimental lower bound. Another advantage
of non-renormalizable interactions is that we get easily an extra mechanism
for CP breaking to fuel baryogenesis, in addition to the usual breaking via the
CKM matrix.

\section{The phase transition}
The effective Higgs potential is the crucial factor for studying the dynamics of
the electroweak phase transition (EWPT). In the high temperature expansion we
get
\begin{eqnarray}\label{Veff}
V_{\rm eff}(\phi,T)=&&{1\over 2}\left(-\mu^2+\left({1\over 2}\lambda+{3\over
      16}g_2^2+{1\over 16}g_1^2+{1\over 4}y _ t ^2\right)T^2\right)
\phi^2 \nonumber\\
&-&{{g^3_2}\over{16\pi}}T\phi^3 +{\lambda \over
  4}\phi^4+{3\over{64\pi^2}}y _ t ^4 \phi^4 \ln \left({Q^2 \over {c_F
      T^2}}\right)\nonumber\\
&+&{1\over{8M^2}}(\phi^6+2\phi^4 T^2+ \phi^2 T^4)
\end{eqnarray}
which includes the thermal mass term, the resummed one-loop contributions due to
the transverse gauge bosons and the top quark as well as the leading one-loop
and two-loop corrections due to the $\phi^6$ interaction. ($y _ t $ is the top
Yukawa coupling, $g_2$ and $g_1$ are the $SU(2)_L$ and $U(1)_Y$ gauge couplings,
$c_F\approx13.94$ and we have chosen $Q=m_{\rm top}=178$ GeV.) It is possible to
express the two unknown parameters $\mu$ and $\lambda$ by the physical
quantities $m_H$ and $v=246$ GeV, so that we treat the Higgs mass $m_H$ and the
cut-off scale $M$ as the free parameters of our model with the constraint
$m_H>114$ GeV. In addition we have to require $M<1$ TeV to be relevant at weak
scale temperatures and to be consistent with electroweak precision data we
should certainly have $M$ bigger than a few hundred GeV.\\
To compute the strength of the PT we need the ratio of the non-zero value of
the vacuum expectation value $v_c$ and the critical temperature $T_c$. To
determine these two quantities the two conditions
\begin{equation}
\left.{\partial V_{\rm eff}(\phi,T_c)\over \partial \phi}\right|_{\phi=v_c}=0
\mbox{\hspace{1.0cm}and\hspace{1.0cm}} V_{\rm eff}(v_c,T_c)=0
\end{equation}
have to be fulfilled. Typically $T_c$ is around 100 GeV in case of the EWPT.\\
In
\begin{figure}
\begin{center} 
   \epsfig{file=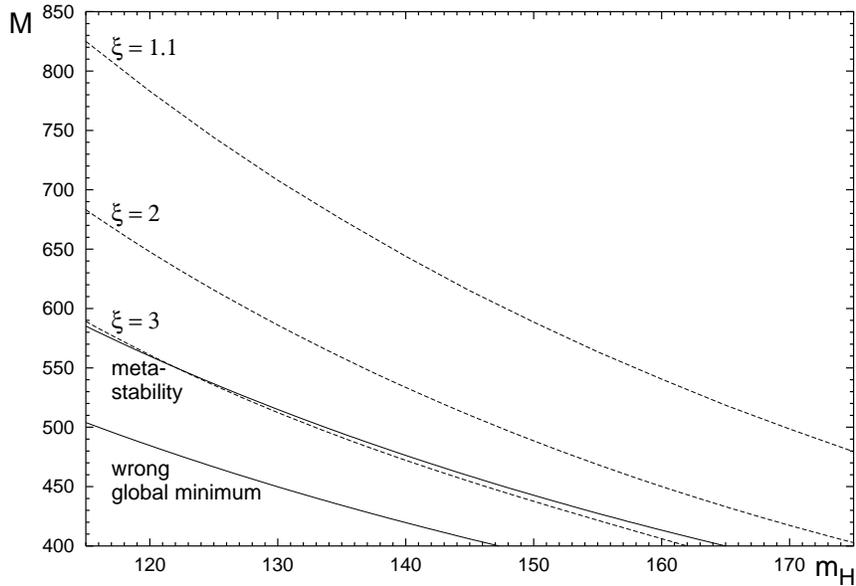,width=80mm,angle=270}
\end{center}
\caption{Contours of constant $\xi=v_c/T_c$ in the $M$-$m_H$-plane. $M$ and $m_H$
         are given in units of GeV.}
\label{figure1}
\end{figure}
fig.~\ref{figure1} the strength of the PT is shown as a function of the model
parameters. It becomes weaker for increasing Higgs masses and for the smallest
allowed Higgs mass we need $M\lsim$ 825 GeV to satisfy the washout
criterion. Below the lowest line (indicated by "wrong global minimum") the
symmetric vacuum is the global one for every temperature meaning the non-zero
vacuum is metastable and there is no longer a PT. Below the "metastability" line
the probability for thermal tunneling gets too small compared to the Hubble
expansion rate, that means in this region the universe remains stuck in the
symmetric vacuum. But there is a large part of the parameter space satisfying
the necessary condition for electroweak baryogenesis to succeed. In contrast
to the SM we find a strongly first order phase transition, even for 
Higgs masses up to 170 GeV.

\section{Bubble characteristics}
The wall thickness $L_{\rm w}$ and the velocity $v_{\rm w}$ are the bubble
properties which will enter the following computation of the baryon asymmetry.\\
What do we mean by the wall thickness and how to estimate it? The solution
of the field equation can be approximately described by a kink,
\begin{equation}
\phi(z)={v_c\over 2}\left(1-\mbox{tanh}{z\over L_{\rm w}}\right).
\end{equation}
We could show that this kink solution using the estimate
$L_{\rm w}=\sqrt{v_c^2/(8V_b)}$, where $V_b$ is the height of the potential
barrier, fits the wall profile quite well. $L_{\rm w}$ varies in a wide range
between 2 and $14T_c$ depending on the combination of the parameters. It becomes
smaller as we decrease $m_H$ at fixed $M$ and it is the same the other way round
if we decrease $M$ at fixed $m_H$.
\begin{figure}
\begin{center}
   \epsfig{file=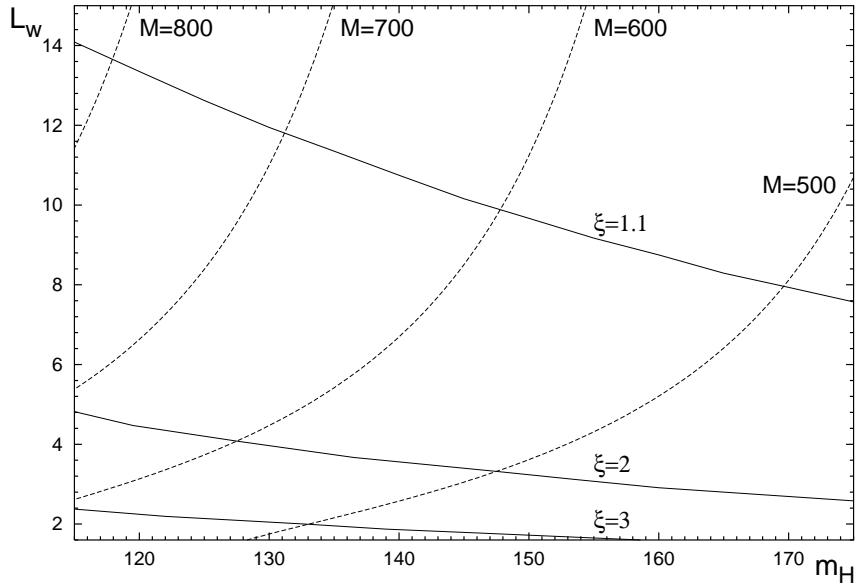,width=80mm,angle=270}
\end{center}
\caption{The wall thickness $L_{\rm w}$ in units of $T_c$ as a function of the Higgs mass $m_H$ for
         several values of $M$. $m_H$ and $M$ are given in units of GeV.}
\label{figure2}
\end{figure}
In fig. \ref{figure2} this behaviour is presented in
dependence of the model parameters. In addition lines of constant $\xi$ are
shown.\\
The computation of the wall velocity is very complex. On the one
hand the expansion of the bubbles is accelerated by the internal pressure and on
the other hand it is slowed down by plasma friction. In addition $v_{\rm w}$ is
also reduced by latent heat of the nucleating bubbles. The wall propagates with a
constant velocity when the forces are balanced and a stationary
situation is reached. In general, the movement of the wall is faster in the case of a
stronger PT. But it is only possible to give rough estimates with large
uncertainties so that we decided to treat $v_{\rm w}$ as a free parameter in the
following computation of the baryon asymmetry.

\section{CP violation}
A new source of CP violation is generated by introducing a dimension-6
Higgs-fermion interaction with the Lagrangian
\begin{equation}
{\cal{L}}_m=\bar{\Psi}_L\left(y\Phi+{x\over
    M^2}\left(\Phi^{\dagger}\Phi\right)\Phi\right)\Psi_R+h.c.
\end{equation}
where $y$ stands for the usual Yukawa couplings and $x$ for the new
couplings containing complex phases relative to $y$. In general, these $x$ are
of unknown flavour structure but we assume that they have the same structure as
the corresponding Yukawa couplings. Then we can concentrate on the top quark and
choose for simplification $y_t$ real and $x_t$ purely imaginary. (This is a good
approximation since the real part of $x_t$ is negligible in comparison to $y_t$
because of the $M^2$ in the denominator. So only its imaginary
part is of relevance.) The Lagrangian then reduces to
\begin{equation}
{\cal{L}}_m=\bar{t}_Lm_t t_R+\bar{t}_Rm_t^* t_L
\end{equation}
with the complex top mass
\begin{equation}
m_t=y_t\frac{\phi(z)}{\sqrt{2}}+i|x_t|\frac{{\phi(z)}^3}{2\sqrt{2}M^2}=m(z)e^{i\theta(z)}
\end{equation}
where we have defined a CP violating phase
\begin{equation}\label{theta}
\theta(z)=\arctan\left(\frac{|x_t|{\phi(z)}^2}{2y_t
    M^2}\right).
\end{equation}
Obviously this is C conserving but P violating. In the WKB approach this leads
to different dispersion relations for particles and anti-particles, depending on
their complex mass. These dispersion relations induce different force terms in
the transport equations which describe the evolution of the plasma. So the CP
violating interactions create an excess of left-handed quarks over the
corresponding anti-quarks. Sphaleron transitions where only left-handed
particles are involved annihilate this asymmetry in the symmetric phase. The
expanding bubble sweeps over this region and in the broken phase these
(B+L)-violating reactions are suppressed because of the strongly first order
PT. As a result a net baryon asymmetry is generated.

\section{The evolution of the plasma}
We model the evolution of the particle distributions $f_i(t,{\bf x},{\bf p})$
by classical Boltzmann equations
\begin{equation}
\left(\partial_t+\dot{\bf x}\partial_{\bf x}+\dot{\bf p}\partial_{\bf p}\right)
    f_i(t,{\bf x},{\bf p})={\cal{C}}[f_i]. 
\end{equation}
The index $i$ stands for the type of the particle and $\cal{C}$ is the collision
term. The dispersion relations come into play through
\begin{equation}
\dot{\bf x}=\partial_{\bf p}E_i({\bf x},{\bf p}),\hspace{1.5cm}\dot{\bf
  p}=-\partial_{\bf x}E_i({\bf x},{\bf p}).
\end{equation}
Using the fluid-type ansatz in the rest frame of the plasma \cite{JPT95}
\begin{equation}
  f_i(t,{\bf x},{\bf p})=\frac{1}{e^{\beta(E_i-v_ip_z-\mu_i)}\pm1},
\end{equation} 
we introduce velocity perturbations $v_i$ and chemical potentials $\mu_i$ for each
fluid. The velocity perturbations describe the particle movement in response to
the force caused by the different dispersions relations.\\
We search for stationary solutions of these Boltzmann equations by expanding in
derivatives of the fermion mass. If we weigh the Boltzmann equations with 1 and
$p_z$ we derive after momentum averaging to first order in derivatives
\begin{eqnarray}\label{1a}
\kappa_iv_{\rm w}\mu_{i,1}'-K_{1,i} v_{i,1}'-\sum_p \Gamma_p^{\rm inel}\sum_j\mu_{j,1}
&=&K_{3,i} v_{\rm w} (m_i^2)'
\\[.3cm] \label{1b}
-K_{1,i} \mu_{i,1}'+K_{2,i} v_{\rm w} v_{i,1}'-{K^2_{1,i}\over \kappa_i D_i}v_{i,1}&=&0.
\end{eqnarray}
$\mu_{i,1}$ and $v_{i,1}$ denote the perturbations to first order in
derivatives. At this stage we do not have to distinguish between particles and
anti-particles. To simplify the notation the symbols $K_i$ and $\kappa_i$
betoken some mass dependent momentum averages. The statistical factor $\kappa_i$
would be 1 for massless fermions. A prime represents the derivative with respect
to the relative coordinate perpendicular to the wall ($\bar z\equiv z-v_{\rm
  w}t$). Here the collision terms have already been expressed by inelastic rates
$\Gamma_p^{\rm inel}$ for a process $p$ and by diffusion constants
$D_i$. Realize on the right-hand side of the first equation that the change in
the particle mass along the wall causes the force term for the particles. (The
used notation and a detailed computation can be found in \cite{wir}.)\\
To second order in derivatives there is a difference between particles 
and anti-particles. The perturbations contain CP odd and even components but we
are only interested in the CP odd ones. After subtracting the equations of
particles and anti-particles we derive
\begin{eqnarray} \label{2a}
\kappa_iv_{\rm w}\mu_{i,2}'-K_{1,i} v_{i,2}'-\sum_p \Gamma_p^{\rm inel}\sum_j\mu_{j,2}
&=&-K_{6,i} \theta_i'm_i^2\mu_{i,1}'
\\[.3cm] 
-K_{1,i} \mu_{i,2}'
+K_{2,i} v_{\rm w} v_{i,2}'-{K^2_{1,i}\over \kappa_i D_i}v_{i,2}
&=& K_{4,i} v_{\rm w}m_i^2\theta_i'' 
+K_{5,i} v_{\rm w}(m_i^2)'\theta_i'\nonumber\\
&&-K_{7,i} m_i^2\theta_i'v_{i,1}'.
\label{2b}
\end{eqnarray}
Every CP violating source term on the right-hand side is proportional to
derivatives of $\theta_i$. The source terms proportional to the first
order perturbations can enhance the asymmetry in the baryon number. In the
following section we analyze their relevance.\\
In principle we have to solve this set of differential equations (\ref{1a}),
(\ref{1b}), (\ref{2a}) and (\ref{2b}) for every particle type $i$ to compute the
asymmetry in the left-handed quark density. The chemical potential $\mu_{B_L}$
of the left-handed quarks is then given by the sum over the flavours. (For
details see again \cite{wir}.)\\
The weak sphalerons convert $\mu_{B_L}$ into a baryon asymmetry by \cite{CJK00}
\begin{equation} \label{eta1}
\eta_B=\frac{n_B}{s}=\frac{405\bar\Gamma_{ws}}{4\pi^2v_{\rm w}g_*T^4}\int_0^{\infty}
d\bar z \mu_{B_L}(\bar z)e^{-\nu\bar z}.
\end{equation}
Here $\bar\Gamma_{ws}=1.0\cdot10^{-6}T^4$ \cite{Mws} is the weak sphaleron rate,
$g_*=106.75$ the effective number of degrees of freedom in the plasma and
$\nu=45\bar\Gamma_{ws}/(4v_{\rm w}T^3)$.

\section{Numerical results}
\begin{figure}
\begin{center}
   \epsfig{file=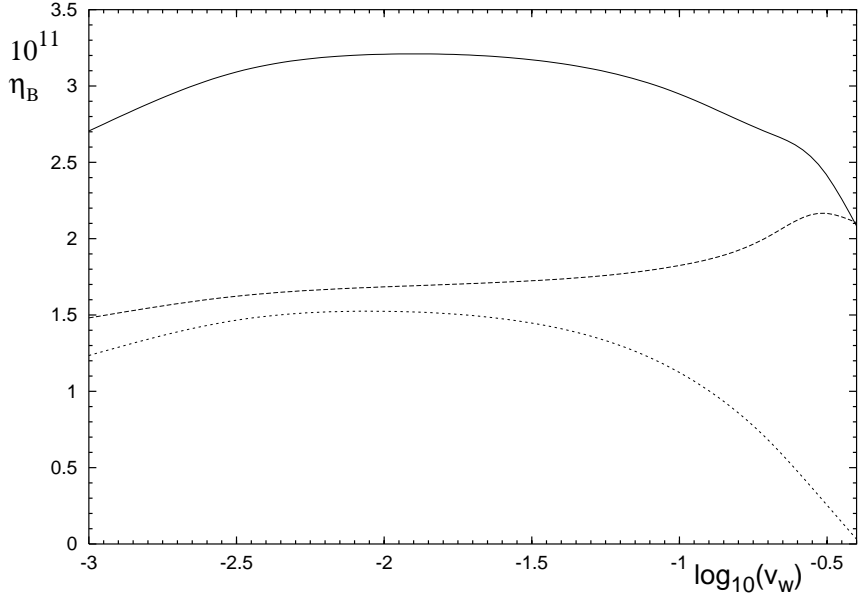,width=80mm,angle=270}
\end{center}
\caption{Comparison of the contributions of the different source terms to the
  total baryon asymmetry as a function of $v_{\rm w}$ for the parameter set
  $L_{\rm w}=8/T$, $M=6T$ and $\xi=1.5$.}
\label{figure3}
\end{figure}

\begin{figure}
\begin{center}
   \epsfig{file=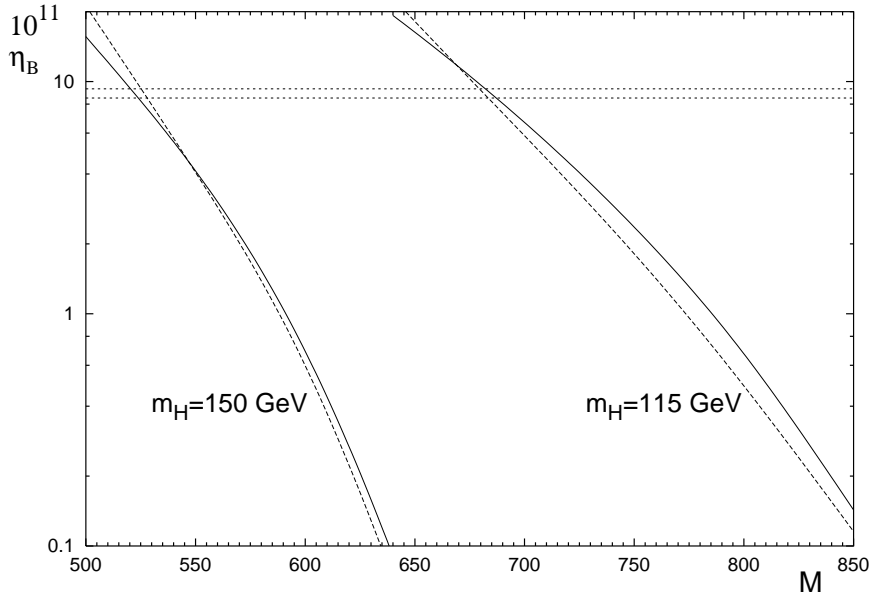,width=80mm,angle=270}
\end{center}
\caption{$\eta_B$ as a function of $M$ (in units of GeV) for two different Higgs
  masses respectively for $v_{\rm w}=0.01$ (solid) and $v_{\rm w}=0.3$
  (dashed).}
\label{figure4}
\end{figure}

In this section we present the results of our computation. The most important
particles to be taken into account are the left- and right-handed top quarks and
the Higgs bosons. We can neglect the other quarks and leptons due to their small
masses. We found that also the Higgs bosons can be ignored because their
influence on the generated baryon asymmetry is negligible. So we have to solve
the equations (\ref{1a}), (\ref{1b}), (\ref{2a}) and (\ref{2b}) simultaneously
only for the left-handed top and bottom quark as well as for the right-handed
top quark to compute the asymmetry in the left-handed quark density. At this
stage we assume that the gauge interactions are in equilibrium. Moreover we
neglect the weak sphalerons, i.e. baryon and lepton number are conserved. We
take into consideration the strong sphalerons $(\Gamma_{ss})$, the top Yukawa
interaction $(\Gamma_y)$ and the top helicity flips $(\Gamma_m)$ which are only
present in the broken phase. In our numerical computation we use the values
$\Gamma_{ss}=4.9\cdot10^{-4}T$ \cite{Mss}, $\Gamma_y=4.2\cdot10^{-3}T$
\cite{HN95}, $\Gamma_m=m_t^2(\bar z,T)/(63T)$ for the different rates and
$D=6/T$ \cite{HN95} for the quark diffusion constant. To compute the thermal
averages we use the half $m_t^2$ of the broken phase.\\
As already mentioned there are different source terms appearing in the transport
equations (\ref{2a}) and (\ref{2b}). In fig.~\ref{figure3} we compare their
contributions to the generated baryon asymmetry for one typical set of
parameters. The
lower line represents the contribution to $\eta_B$ which arises from the source
terms proportional to the first order perturbations $\mu_{i,1}'$ and $v_{i,1}'$. Note
that these source terms are non-negligible, especially for small wall
velocities. The middle line shows the contribution due to the other
sources and the upper line is the total baryon asymmetry,
that means the sum of the two parts. Another general result is that $\eta_B$
depends only slightly on $v_{\rm w}$.\\
We also found that $\eta_B$ decreases with increasing $L_{\rm w}$. As
expected $\eta_B$ increases rapidly for a stronger PT (larger values of $\xi$) since
then the top mass becomes larger and the effect of the CP violating phase $\theta$
stronger. These are general remarks about the influence of the different parameters
to the generated baryon asymmetry. If we look in detail at the model under
consideration, we have to compute the strength of the PT and the bubble wall
width for every value of the cut-off scale $M$. In our computation we take
$|x_t|=1$ for the CP violating phase (\ref{theta}). In fig.~\ref{figure4} we
present $\eta_B$ as a function of $M$ for two different Higgs masses $m_H=115$
and 150 GeV. We have chosen one large $v_{\rm w}=0.3$ and one small wall
velocity $v_{\rm w}=0.01$ to show again the small dependence on this
parameter. The measured value (\ref{eta}) is signified by the horizontal
lines in the figure.
For every Higgs mass the baryon asymmetry gets smaller for increasing $M$. But
we can easily generate the observed value for natural values of the parameters!

\section{Conclusions}
We have investigated the EWPT and baryogenesis in the SM extended by
dimension-6 Higgs interactions. Non-renormalizable operators parameterize new
physics beyond some cut-off scale $M$. This scale and the Higgs mass are the two
free parameters of this model. Using the one-loop thermal potential we find a
large part of the parameter space where the requirements of electroweak
baryogenesis are satisfied. Even for Higgs masses up to 170 GeV the PT is strong
enough to avoid baryon number washout.\\
A dimension-6 Higgs-fermion interaction arranges a new source of CP
violation. We obtain a complex phase in the quark mass which varies along the
bubble wall. This causes different forces on the quarks and corresponding
anti-quarks as they pass through the phase boundary. As a result, an excess of
left-handed quarks is created which afterwards is converted into a baryon
asymmetry by the sphaleron transitions.\\
We describe the evolution of the plasma
in the WKB approach and expand in derivatives of the wall profile. We found
novel source terms in the transport equations. These source terms proportional
to the first order perturbations enhance the generated baryon asymmetry
especially for small wall velocities.\\
The model under consideration provides the missing
ingredients for electroweak baryogenesis, i.e. a strong first order phase
transition and additional CP violation. The measured baryon to entropy ratio can
be explained for natural values of the parameters.

\section*{Acknowledgements}
I would like to thank Dietrich B\"odeker, Stephan J. Huber and Michael Seniuch
for a very fruitful and pleasant collaboration as well as Steffen Weinstock and
J\"org Erdmann for helpful discussions. This work was supported by the DFG,
grant FOR 339/2-1.

\end{document}